\documentclass[aps,pra, twocolumn]{revtex4-1}
\usepackage{amsthm}

\usepackage[utf8]{inputenc}
\usepackage[sc,osf]{mathpazo}

\usepackage{graphicx,subfigure}
\usepackage{bm}
\usepackage{makeidx} 
\usepackage{amsmath}
\usepackage{amssymb}
\usepackage{color}
\usepackage{amsthm}
\usepackage{hyperref}
\usepackage{cleveref}
\usepackage{diagbox}
\hypersetup{colorlinks, citecolor=magenta, filecolor=blue, linkcolor=blue, urlcolor=green}
\usepackage[normalem]{ulem}
\usepackage{float}
\usepackage{multirow}

\newcolumntype{P}[1]{>{\centering\arraybackslash}p{#1}}
\begin{document}

\title{Efficient nonlinear witnessing of non-absolutely separable states  with lossy detectors}

\author{Ayan Patra$^{1}$, Shiladitya Mal$^{1, 2}$, Aditi Sen(De)$^{1}$}
	
	\affiliation{$^1$ Harish-Chandra Research Institute and HBNI, Chhatnag Road, Jhunsi, Allahabad - 211019, India\\
	$^2$ Department of Physics and Center for Quantum Frontiers of Research and Technology (QFort),
National Cheng Kung University, Tainan 701, Taiwan
	}

\begin{abstract}

Entangled states are undoubtedly an integral part of various quantum information processing tasks. On the other hand, 
 absolutely separable states  which cannot be made entangled under any global unitary operations are useless from the resource theoretic perspective, and hence identifying  non-absolutely separable states can be an important issue for designing quantum technologies.  Here we report that nonlinear witness operators provide significant improvements in detecting non-absolutely separable states over their linear analogs, by invoking examples of states in various dimensions. We also address the problem of closing detection loophole and find critical efficiency of detectors above which no fake detection of non-absolutely separable (non-absolutely positive partial transposed) states is possible.

\end{abstract}

\maketitle

\section{Introduction}
\label{sec:intro}

With the advent of quantum information science, shared entanglement \cite{horodecki'09} turns out to be one of the main resource for quantum technologies \cite{nielsen} which include quantum key distribution \cite{ekert'91}, dense coding \cite{dc}, teleportation \cite{tele}, clock synchronisation \cite{clock}, remote state preparation \cite{rsp}, measurement-based computation \cite{computation} to name a few. In the theory of entanglement, developing efficient methods for the generation, detection and preservation of entangled states is an important enterprise \cite{horodecki'09, guhne'09, bookchap} while finding unprofitable resources like  unentangled or separable states  which  can be decomposed in terms of a convex combination of pure product  states \cite{werner'89} also plays a crucial role.
Over the years, several criteria have been proposed for identifying entangled states although according to the computational complexity class, the so-called entanglement-separability problem is NP-hard \cite{gurvits'03, ghariban'10}.  A prominent mathematical detection method is the partial transposition map, a necessary and sufficient criterion for a bipartite system having dimensions up to six,  which is through the existence of all positive but not completely positive maps \cite{peres'96, horo'96}.

Apart from these theoretical endeavors, entanglement witness (EW) provides a powerful tool in the domain of identification of entangled states.  Since the set of the separable state is convex and compact,  the Hahn Banach theorem  ensures that there exists a witness operator for each entangled states \cite{horo'96, terhal'20}. EWs are Hermitian operators, whose   expectation value with respect to all separable states is non-negative while it gives negative value for at least one entangled state. Importantly, they provide an efficient method of detecting entanglement in laboratories via local measurements, thereby establishing themselves as useful \cite{guhne'02}.

As far as production of entanglement is concerned, separable states can be made entangled by suitable joint unitary operations \cite{horodecki'09}. On the contrary, there exists a class of separable states which cannot be made entangled by the application of any joint unitary gate,  known as absolutely separable (AS) states or separable states from spectrum \cite{knill'03, kusz'01, vers'01, hilde'07, slater'09, johnston'13, nirman'14, aruna'15, jivu'15}.  In a similar fashion, absolutely positive partial transposed (PPT) states are introduced. Although beyond qubit-qutrit states, PPT and separability are not equivalent due to the existence of PPT bound entangled states \cite{horostate'97, horo'99},  interestingly, it was found that  absolute-PPT and AS are same in these dimensions  \cite{hilde'07, johnston'13}. From the perspective of resource theory,  AS states or absolute-PPT state are a kind of free or useless states and hence detecting them are significant to identify resources. In the recent past,  witnesses for non-AS states have been proposed in a similar spirit of linear EWs  \cite{nirman'14}. 

In the domain of detecting entanglement, another interesting twist  comes from the possibility of improving linear witness operators by finding the way of obtaining nonlinear EWs \cite{guhne'06, guhne'07, zhang'07,  moroder'08,  koto'10, nl0, nl1, nl2, ks'19, nl3}. It was shown that every EW can be upgraded by adding nonlinear term(s) and the novel technique enables us to show that the set of separable states cannot have facets \cite{guhne'07}, thereby shed light in the geometry of quantum states \cite{beng'06} (see recent results on the boundary of the set of AS states  \cite{jivu'15a, hmd}).  

In the present work, we explore nonlinear improvement of linear witnesses for determining non-AS states. To this end, some expectation values in quadratic form have to be subtracted from the linear witness operators by maintaining their essential properties. We consider two ways of modifying linear witnesses \cite{guhne'06} -- one is by subtracting a single while  in the other case,  terms corresponding to a full basis set are subtracted. The method is demonstrated by considering a class of two-qubit, qubit-qudit, and two-qutrit states. In all these cases, we show that nonlinear witnesses perform better for detecting non-AS or non-absolutely PPT states in comparison with the linear ones. Specifically, in higher dimensions, we propose a class of absolutely separable as well as absolutely PPT states by mixing bound entangled states with white noise and construct nonlinear entanglement witnesses explicitly to detect non-absolutely PPT states. Towards obtaining the results, we also provide  classes of  global unitary operators which can transform PPT states to the one having non-positive partial transposition (NPPT).


Finally, we discuss the change in behavior of these nonlinear witnesses for non-AS states in a more realistic condition i.e., under  inefficient detectors. Specifically, we consider the scenario when the  detectors may not click indicating lost events. Addressing detection loophole in Bell test is an old problem \cite{btdl}. In the context of EW,  critical detector efficiency above which no fake detection of entanglement is possible was derived in Ref. \cite{dl'07}. Recently similar conditions for closing detection loophole in the context of nonlinear  and measurement device-independent EWs are reported \cite{ks'19, mdl'21}. In the context of identification of non-AS states, we also show  that nonlinear witnesses provide substantial improvements in critical detector efficiencies than that obtained via linear witness operators.

We organize the paper in the following way. In Sec. \ref{sec:prelim}, the mathematical condition for absolutely separable as well as absolutely PPT states and nonlinear entanglement witness operators are presented. To demonstrate the power of nonlinear witnesses, a class of two-qubit non-AS states is considered in Sec. \ref{sec:twoqubits} while the classes of non-absolutely PPT states are constructed and their identification methods via nonlinear witnesses are exhibited in Sec. \ref{sec:highdim}. In the next section (Sec. \ref{sec:loophole}), we provide a method to overcome inefficiencies in detection procedure and finally we conclude in Sec. \ref{sec:conclu}. 

 \section{Preliminary concepts, definitions and notations}
 \label{sec:prelim}
 
Before going into the main results, we first present the known  criteria for detecting absolutely separable and PPT states. Let us also briefly introduce the nonlinear witness operators for the set of  non-absolutely  separable as well as non-absolutely PPT states.    

\subsection{ Criteria for absolutely separable and positive partial transpose  states and their witness operators}

A subset of separable states, which can not be made entangled using any global unitary operation are called absolutely separable states and similarly absolutely PPT state are those which remains PPT even after applying any global unitary operation.

If a bipartite state $\rho_{AB}$ in $2\otimes n$ dimension \cite{dimexplain}  has eigenvalues $\lambda_{1}$,$\lambda_{2}, \cdots, \lambda_{2n}$ in descending order,  the condition for the absolute separability reads as \cite{hilde'07}
\begin{equation}
\label{2n}
\lambda_{1}-\lambda_{2n-1}-2\sqrt{\lambda_{2n-2}\lambda_{2n}}\leq 0.
\end{equation}
Interestingly, note that in \(2 \otimes n\), the set containing absolutely PPT states coincides with the set of absolutely separable states \cite{johnston'13} which is not true for the partial transposition criteria in the entanglement-separability paradigm for \(n>3\). 

On the other hand, in higher dimensions, i.e., in $3 \otimes n$,  let the eigenvalues of $\rho_{AB}$ be $\lambda_{1}$,$\lambda_{2},\cdots,\lambda_{3n}$ organized in the descending order. The states are  absolutely PPT  \cite{hilde'07}, when they satisfy the following conditions:
\begin{equation}
\label{eq:absPPT1}
\begin{split}
\begin{vmatrix} 2\lambda_{3n} & \lambda_{3n-1}-\lambda_{1} & \lambda_{3n-3}-\lambda_{2} \\ \lambda_{3n-1}-\lambda_{1} & 2\lambda_{3n-2} & \lambda_{3n-4}-\lambda_{3} \\ \lambda_{3n-3}-\lambda_{2} & \lambda_{3n-4}-\lambda_{3} & 2\lambda_{3n-5}\end{vmatrix} \geq 0,
\end{split}
\end{equation}
and 
\begin{equation}
\label{eq:absPPT2}
\begin{split}
\begin{vmatrix} 2\lambda_{3n} & \lambda_{3n-1}-\lambda_{1} & \lambda_{3n-2}-\lambda_{2} \\ \lambda_{3n-1}-\lambda_{1} & 2\lambda_{3n-3} & \lambda_{3n-4}-\lambda_{3} \\ \lambda_{3n-2}-\lambda_{2} & \lambda_{3n-4}-\lambda_{3} & 2\lambda_{3n-5}\end{vmatrix} \geq 0.
\end{split}
\end{equation}
Notice that there exists no simple criteria for absolutely separable states in higher dimensions. Moreover, with the increase of dimensions, finding all the eigenvalues for checking the above criteria requires full tomography \cite{tomo} which becomes cumbersome.  Hence the witness operators can play a crucial role to detect non-absolutely PPT states, both theoretically and experimentally.

\subsection{Nonlinear witnesses for non-absolutely PPT states}

In functional analysis,  a celebrated theorem, known as Hahn Banach separation theorem, states that if $S_{1}$ and $S_{2}$ be two nonempty, convex disjoint subsets of a normed linear space $V$,  there is a hyperplane that separates $S_{1}$ and $S_{2}$ \cite{func}. Moreover, if $A$ and $B$ are two nonempty disjoint subsets  of a normed linear vector space $V$, where one of them, say, $A$ is convex,  there exists a hyperplane, serving as an witness operator which can separate the entire subset $A$ from any point of $B$.  
Since separable states as well as absolutely separable states both form compact and convex sets,  a Hermitian operator, $W$  for which $\mbox{Tr}(\sigma W)\geq 0$ for all separable states (absolutely separable states), $\sigma$, and $\mbox{Tr}(\rho W) < 0$ for at least one entangled state (non-AS state), $\rho$, is called the linear witness operator. 
They are linear witnesses since   linear expression of mean values of $W$ is involved in the definition. Note also that the set of absolutely PPT states is also compact, and hence one can construct witness operators which can detect non-absolutely PPT states. 
Given two EWs, $W_1$ and $W_2$,  $W_2$ is said to be finer than $W_1$ if it  can detect all states which are also identified by $W_1$ while  an EW is called optimal if there is no other witness finer than it \cite{oew}.

On the other hand,  it was shown that the  linear witness operators can  always be upgraded  according to its capability of identifying non-separable states by introducing nonlinearity  \cite{guhne'06, guhne'07}. 
For example, let us consider states whose entanglement can be detected by NPPT.  In this case, non-separability can be witnessed by $|\phi\rangle\langle\phi|^{T_B}$ where $|\phi\rangle$ is the eigenvector corresponding to the negative eigenvalue of $\rho^{T_B}$.  As prescribed in \cite{guhne'06}, one can introduce nonlinearity in the following ways.
\begin{equation}\label{f1}
F^1(\rho)=\langle|\phi\rangle\langle\phi|^{T_B}\rangle-\dfrac{1}{S(\psi)}\langle{X}^{T_B}\rangle\langle({X}^{T_B})^\dagger\rangle
\end{equation}
and 
\begin{equation}\label{f2}
F^2(\rho)=\langle|\phi\rangle\langle\phi|^{T_B}\rangle-\sum_{i=1}^{k}\langle{X}_i^{T_B}\rangle\langle({X}_i^{T_B})^\dagger\rangle
\end{equation}
where all the expectation values are taken with respect to the given state $\rho$. Here in Eq. (\ref{f1}), $X$ is given by $|\phi\rangle\langle\psi|$, where $|\psi\rangle$ is any arbitrary state and $S(\psi)$ is the square of the largest Schmidt coefficient of the state $|\psi\rangle$ while in Eq. (\ref{f2}), $X_i$ is defined by $|\phi\rangle\langle\psi_i|$ ($i=1,2, \cdots, d$) with an orthonormal basis being $\{|\psi_i\rangle\}$.

Let us now adopt the similar procedure to detect 
 non-absolutely separable (non-absolutely PPT) states, say, $\rho'$ which can be made entangled from  a separable (PPT) one by global unitary operator $U$
 \cite{nirman'14}. In this situation, to detect \(\rho'\), the witness operator, $|\phi\rangle\langle\phi|^{T_B}$ should be modified as 
 $U^\dagger|\phi\rangle\langle\phi|^{T_B}U$,  where $|\phi\rangle$ is the eigenvector corresponding to the negative eigenvalue of $({U}\rho'{U}^{\dagger})^{T_B}$. 
 
 Let us now   incorporate the nonlinear terms into  witness operators for determining non-AS (non-PPT) states.   Like  Eqs. (\ref{f1}) and (\ref{f2}), the detection method for non-AS (non-absolutely PPT) states take the form as
\begin{eqnarray}
\label{eqmodnlW}
 F^1(\rho')& =& \langle{U}^\dagger|\phi\rangle\langle\phi|^{T_B}U\rangle-\dfrac{1}{S(\psi)}\langle{U}^\dagger{X}^{T_B}U\rangle\langle(U^\dagger{X}^{T_B}U)^\dagger\rangle,  \nonumber\\
 F^2(\rho') &=&\langle{U}^\dagger|\phi\rangle\langle\phi|^{T_B}U\rangle \nonumber \\
& - &\sum_{i=1}^{k}\langle{U}^\dagger{X_i}^{T_B}U\rangle\langle(U^\dagger{X_i}^{T_B}U)^\dagger\rangle,
\end{eqnarray}
where $X^{T_B}$  is replaced by $U^\dagger{X}^{T_B}U$ and all the expectation values have to be taken with respect to the given state $\rho'$. As we will show that in the detection process of non-absolutely separable  (PPT) states,  finding a nontrivial global $U$  is, in general, difficult.



\begin{figure}[ht] 
\includegraphics[width=0.9\linewidth]{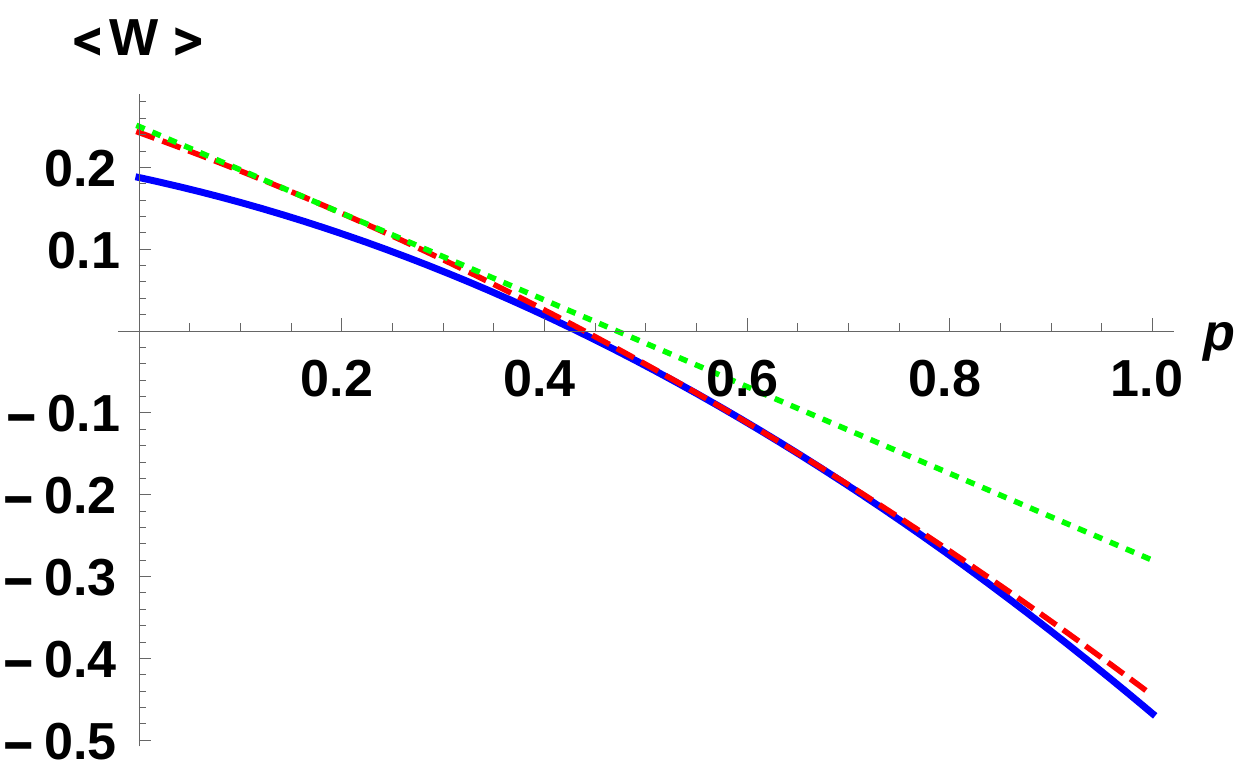}
\caption{(Color online.) Witness operators, \(\langle W \rangle\),  vs. $p$ of the generalized Werner state, \(\rho_{gW}\) in Eq. (\ref{eq:rhogW}). 
Solid (blue),  dashed (red) and  dotted (green) lines correspond to nonlinear witness operators, $F^2$, $F^1$ and linear witnesses respectively. 
Here $\alpha=\pi/12$ in \(|\xi \rangle\)  and $\theta=0.352$ involved in the witness operator, \(|\phi\rangle\), in radians. Both the axes are dimensionless. 
}
\label{fig:1}
\end{figure}

\section{Nonlinear improvement in witnessing two-qubit  non-absolutely separable states}
\label{sec:twoqubits} 

 We will now explicitly show advantages of nonlinear witness over their linear counterparts by constructing them and by invoking a class of two-qubit states.

We now use the criteria, given in  (\ref{2n}) to find the range of the state parameter for which the state is absolutely separable. To illustrate the method, let us consider a generalized Werner state \cite{werner'89},
\begin{equation}
\label{eq:rhogW}
\rho_{gW} =p |\xi\rangle\langle\xi|+\dfrac{1- p }{4} I_{4\times 4}.
\end{equation}
where $|\xi\rangle= \cos \alpha |00\rangle+ e^{i \phi} \sin \alpha|11\rangle$, \(p\) is the mixing parameter and \(I_{4 \times 4}\) is the identity operator with the subscript representing the size of the matrix, thereby indicating the dimension of the system. 
Following the condition written in (\ref{2n}) with $2n=4$, we find that 
 $\rho_{gW}$ is absolutely separable  in the range $0\leq p \leq\frac{1}{3}$ which is independent of \(\alpha\) and \(\phi\), while the state is entangled when  $\dfrac{1}{1+2 \sin2\alpha} < p \leq 1 $. 
Except \(\alpha= \pi/4\), there exists a range of \(p\) in which the state is separable but not absolutely separable.   For example, for $\alpha=\frac{\pi}{12}$, i.e., $\sin 2\alpha=\frac{1}{2}$,  the state $\rho_{gW}$  is entangled with $\dfrac{1}{2}< p \leq 1$ and separable but not absolutely separable in \(1/3 < p \leq 1/2\). It implies that there exists a global unitary operator which  
can convert the state $\rho_{gW}$ to an entangled state (or, NPPT in  $2\otimes2$ ), say $\rho_e$ in the range $\frac{1}{3}< p \leq\frac{1}{1+2 \sin2\alpha}$. 

Let us consider a global unitary operator, given by
\begin{eqnarray}
U=\dfrac{1}{\sqrt{2}}
\begin{pmatrix}
1 & 0 & 0 & 1 \\ 0 & \sqrt{2} & 0 & 0 \\ 0 & 0 & \sqrt{2} & 0 \\ -1 & 0 & 0 & 1
\end{pmatrix}.
\end{eqnarray}
 After applying this unitary operator,   $\rho_{gW}$ becomes $\rho_e$   for $p > 0.366$, i.e. when $0.366 < p <0.5$, the state is entangled although it  was initially separable, i.e.,  PPT.

By choosing  $|\phi\rangle =\frac{1}{\sqrt{2}}(|01\rangle+|10\rangle)$, we can show that $U^\dagger|\phi\rangle\langle\phi|^{T_B}U$ can detect the $\rho_{gW}$ as non-absolutely separable in the range $0.366<p<0.5$.
Let us take $|\phi\rangle = \cos \theta|01\rangle+\sin \theta|10\rangle$, \((0 \leq \theta \leq \pi)\) and in that case, $U^\dagger|\phi\rangle\langle\phi|^{T_B}U$ can detect the state $\rho_{gW}$ as non-absolutely separable in the range $\dfrac{1}{1+\sqrt{3}\sin 2\theta}< p <0.5$, with the condition being
\begin {equation}
\langle W(\rho_1)\rangle=\frac{-\sqrt{3} p \sin2\theta- p +1}{4} < 0.
\end{equation}

To make a relatively weaker witness than optimal with $\theta =\pi/4$, consider any value of $\theta$
between $\pi/10.21$ and $\pi/2.4872$, to detect the range for which the state $\rho_{gW}$ is non-absolutely separable.
Let us now introduce nonlinearity to improve the range of detection. By considering $|\psi\rangle= |01\rangle$, and by using Eq. (\ref{eqmodnlW}),
we can reach to the condition for the state $\rho_{gW}$ to be non-absolutely separable if
\begin{equation}
\begin{aligned}
F^{1}_{|01\rangle}(\rho_1)=\frac{-\sqrt{3} p \sin2\theta- p +1}{4}\\
-\frac{(\cos\theta(1- p )-\sqrt{3} p \sin\theta)^2}{16}<0,
\end{aligned}
\end{equation}
while using $|\psi\rangle$=$|10\rangle$, the condition gets modified as
\begin{equation}
\begin{aligned}
F^{1}_{|10\rangle}(\rho_1)=\frac{-\sqrt{3} p \sin2\theta- p +1}{4}\\
- \frac{(\sin\theta(1- p )-\sqrt{3} p \cos\theta)^2}{16}<0.
\end{aligned}
\end{equation}
Instead of product states, if we  use entangled state as   $|\psi\rangle =\frac{1}{\sqrt{2}}(|01\rangle\pm|10\rangle)$,  we cannot provide any advantage except for some range of $\theta$.
On the other hand,
for obtaining \(F^2\), we choose  an orthonormal basis, 
$\{|\psi\rangle_i= |00\rangle,|01\rangle,|10\rangle,|11\rangle$\}, and obtain the criteria for witnessing non-absolutely separable state as
\begin{equation}
\begin{aligned}
F^{2}(\rho_1)=\dfrac{-\sqrt{3} p \sin 2\theta- p +1}{4}\\-\dfrac{(\cos\theta(1-p )-\sqrt{3}p \sin\theta)^2}{16}\\-\dfrac{(\sin\theta(1- p )-\sqrt{3}p \cos\theta)^2}{16}<0.
\end{aligned}
\label{eq:twoqubitF2}
\end{equation}
Notice that instead of the computational basis, \(\{|\psi_i\rangle \}\), if we take the Bell basis (i.e., entangled  states as basis elements), we reach to  the same condition as above. Fig. \ref{fig:1} depicts the comparison between linear and two kinds of nonlinear witness operators in case of detecting non-absolutely separable state, \(\rho_{gW}\) for fixed  values of \(\alpha\) and \(\theta\).  

\section{Construction of Non-absolutely PPT states  with the conjunction of PPT bound entangled states and their witnesses}
\label{sec:highdim}

In this section, we show the usefulness of nonlinearity in witness 
operators for recognizing non-absolutely separable as well as PPT states in higher dimensions. To illustrate this, we construct absolutely separable (PPT) states by mixing white noise with PPT bound entangled states in \(2 \otimes 4\) and \(3 \otimes 3\). Such examples also shed light on the boundaries of the set of PPT bound entangled states and absolutely PPT states.

\begin{figure}[ht] 
\includegraphics[width=0.9\linewidth]{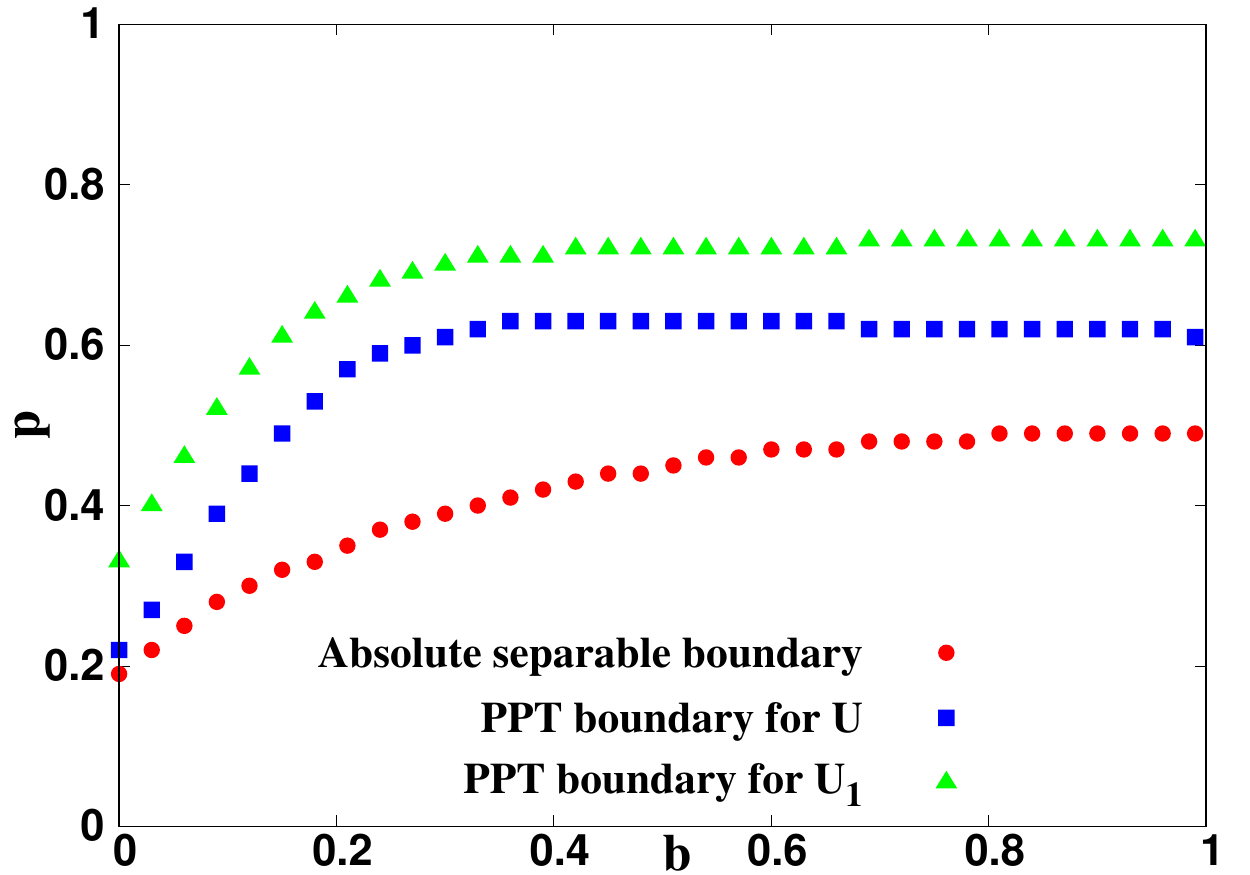}
\caption{(Color online.) Mixing parameter, $p$  against $b$ in \(\rho_2 = p\rho_{b} +\dfrac{1-p}{8}I_{8\times8} \) in Eq. (\ref{eq:rho2}). Circles (red), squares (blue) and triangles (green) represent the boundaries of absolutely separable states and non-AS states, PPT states and NPPT states after applying unitary operators \(U(\pi/3, \pi) \) and \(U_1\) in Eqs. (\ref{eq:U2x4}) and (\ref{eq1:U2x4} respectively. Both the axes are dimensionless.   }
\label{fig:2}
\end{figure}

\subsubsection{Witnessing non-absolutely separable states via PPT bound entangled states}

 In $2\otimes4$ system, let us concentrate on the mixture of the  PPT bound entangled state with white noise, given by
\begin{equation}
\label{eq:rho2}
\rho_{2}=p\rho_{b} +\dfrac{1-p}{8}I_{8\times8},
\end{equation}
where $\rho_b$ is the bound entangled state \cite{horostate'97} represented as
\begin{equation}
\rho_{b}=\dfrac{1}{7b+1}
\begin{pmatrix}
	b & 0 & 0 & 0 & 0 & b & 0 & 0\\ 0 & b & 0 & 0 & 0 & 0 & b & 0 \\ 0 & 0 & b & 0 & 0 & 0 & 0 & b \\ 0 & 0 & 0 & b & 0 & 0 & 0 & 0 \\ 0 & 0 & 0 & 0 & \dfrac{1+b}{2} & 0 & 0 & \dfrac{\sqrt{1-b^2}}{2} \\ b & 0 & 0 & 0 & 0 & b & 0 & 0 \\  0 & b & 0 & 0 & 0 & 0 & b & 0 \\ 0 & 0 & b & 0 & \dfrac{\sqrt{1-b^2}}{2} & 0 & 0 & \dfrac{1+b}{2} \end{pmatrix},
	\end{equation}
with $b\in[0,1]$. Except for $b=0,1$ where $\rho_b$ is separable,  it is PPT bound entangled.
The condition for this state to be absolutely separable, thereby  absolutely PPT in this case is given by (following  inequality (\ref{2n}) with $2n=8$),   
$\lambda_{1}-\lambda_{7}-2\sqrt{\lambda_{6}\lambda_{8}}\leq0$, where $\lambda_i$s are eigenvalues of $\rho_{2}$ in descending order. In Fig. 
\ref{fig:2},  circles indicates the boundary of absolutely separable and non-AS states   in the \((b,p)\)-plane and hence the region below the boundary represents  the absolute separability or absolute PPT of  $\rho_{2}$. Moreover, we find that $\rho_2$ is always PPT for any values of b and p and  therefore, it is important to identify the states lying above the boundary are non-AS states which can be converted to NPPT states with the help of global unitary operators. 
For example, we consider a global unitary operator, $U$, given by
\begin{eqnarray}
\label{eq:U2x4}
   U (\phi_1, \phi_2) &= & a_1\;[\sigma_x\otimes\sigma_y\otimes\sigma_z]+a_2\;[\sigma_y\otimes\sigma_z\otimes\sigma_x] \nonumber \\
 & + &   a_3\;[\sigma_z\otimes\sigma_x\otimes\sigma_y] 
\end{eqnarray}
where $\sigma$'s are  Pauli spin matrices with some parameters $a_1$, $a_2$ and $a_3$, such that $a_1^2+a_2^2+a_3^2=1$. We parametrize them as $a_1=\cos\phi_1,\, a_2=\sin\phi_1\sin\phi_2 \text{ and } a_3=\sin\phi_1\cos\phi_2$, \(0 \leq \phi_1 \leq  \pi, \, 0 \leq \phi_2 \leq 2 \pi\). Taking $\phi_1$ and $\phi_2$ as $\pi/3$ and $\pi$ respectively, we observe that some non-AS states become NPPT which are marked in Fig. \ref{fig:2} by squares. We observe that  there still exists some non-absolute PPT states (lying between the envelopes of circles and squares) which we cannot make NPPT by this unitary operator. 

\begin{figure}[ht] 
\includegraphics[width=0.9\linewidth]{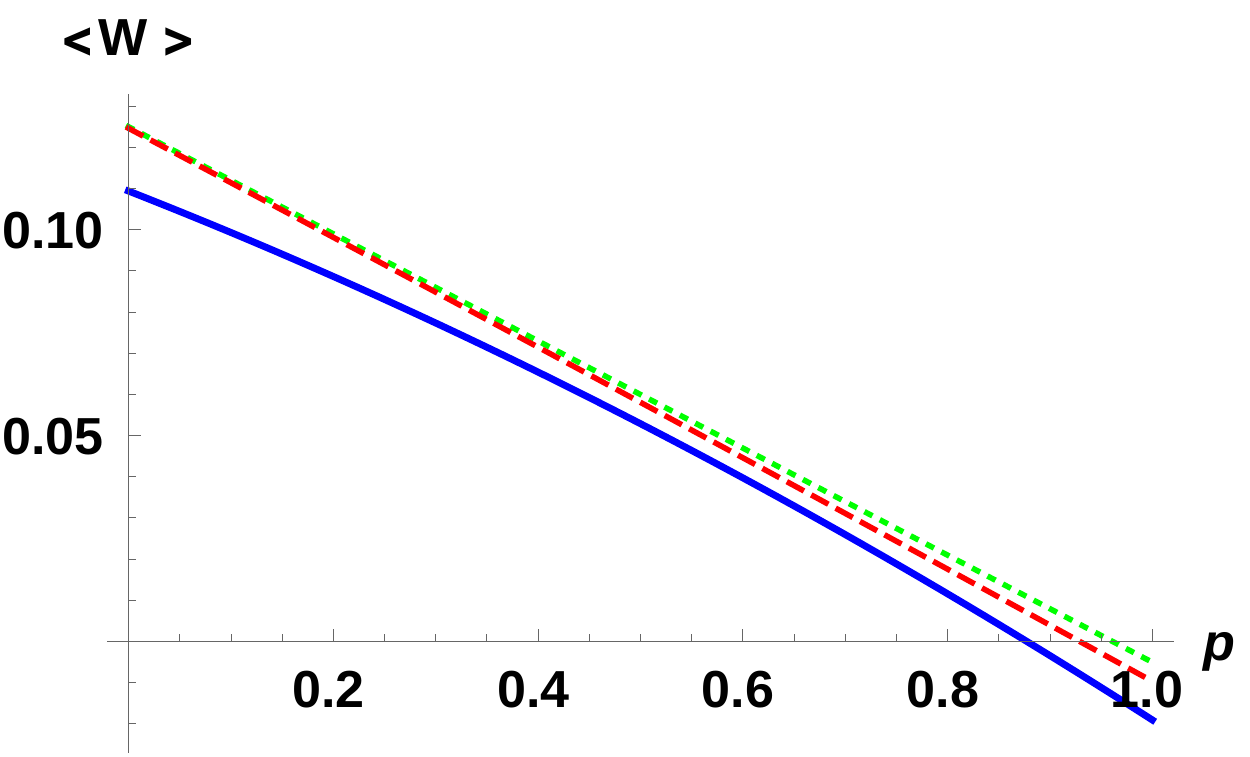}
\caption{Witness operators with \(p\) for \(\rho_2\) in Eq. (\ref{eq:rho2}). Choices of \(\{\theta_i\}\)s in \(|\phi'\rangle\) are mentioned in the text. Here \(b =0.7\). All other specifications are same as in Fig. \ref{fig:1}.  Both the axes are dimensionless.  
}
\label{fig:3}
\end{figure}

At this point, it should be noted that  different values of $b$  require different witnesses to detect the state properly. 
Let us illustrate the entire process for $b=0.7$. 
We find that
 $W=U(\pi/3, \pi)^\dagger|\phi\rangle\langle\phi|^{T_B}U(\pi/3, \pi)$ can detect $\rho_2$ as non-AS state, when $0.62 < p \leq 1$ as shown in Fig. \ref{fig:3}. Here $|\phi\rangle$ is the column vector $(-0.13348,\;0.67743,\;-0.09738,\;0.02271,\;0.00333,\;0.04054,\\-0.71427,\;0.03788)^{T}$ . It is the best linear witness operator for  $b= 0.7$ and for a given $U(\pi/3, \pi)$, since  $W$ constructed by using \(|\phi\rangle\) can detect all the non-AS states when  $ p \in (0.62,  1]$. 
 
 Let us now take a general $|\phi'\rangle$ = $(\sin\theta_1\sin\theta_2\sin\theta_3\sin\theta_4\sin\theta_5\sin\theta_6\sin\theta_7, \\\sin\theta_1\sin\theta_2\sin\theta_3\sin\theta_4\sin\theta_5\sin\theta_6\cos\theta_7,\; \sin\theta_1\sin\theta_2\\\sin\theta_3\sin\theta_4\sin\theta_5\cos\theta_6,\;\sin\theta_1\sin\theta_2\sin\theta_3\sin\theta_4\cos\theta_5,\\\sin\theta_1\sin\theta_2\sin\theta_3\cos\theta_4,\;\sin\theta_1\sin\theta_2\cos\theta_3,\;\sin\theta_1\cos\theta_2,\\\cos\theta_1)^{T}$ with \(0\leq \theta_i \leq \pi, \, \, i=1, \ldots 6\) and \(0 \leq \theta_7 \leq 2 \pi\).   In case of linear witness operator, $U(\pi/3, \pi) ^\dagger|\phi'\rangle\langle\phi'|^{T_B} U(\pi/3, \pi)$ can also detect  $\rho_2$ as non-AS state in some range of \(p\) depending on the parameter values involved in $|\phi\rangle$. However, we can introduce nonlinearity to improve the range of detection. For this purpose,  we consider $|\psi\rangle$=$\frac{1}{\sqrt{2}}(|00\rangle+|10\rangle)$ for $F^1$-type of nonlinear witness operator and an orthonormal basis $\{|00\rangle,|01\rangle,|02\rangle,|03\rangle,|10\rangle,|11\rangle,|12\rangle,|13\rangle\}$ for $F^2$. Now, using Eq. (\ref{eqmodnlW}), and choosing   $\theta_1=2.07345,\;\theta_2=2.36710,\;\theta_3=1.5128,\;\theta_4=1.508,\;\theta_5=1.5382,\;\theta_6=1.7109,\;\theta_7=0.19455$  in $|\phi\rangle$ (where all values are in radians), we observe a clear improvement over linear witness operators (see Fig. \ref{fig:3}). 
 
 \emph{Remark 1.} The similar method can also be applied for other values of $b$.
 
 \emph{Remark 2.} For the clear demonstration of the utility of nonlinear witness operator, we choose a set of \(\{\theta_i\}\). Different values  of  \(\{\theta_i\}\) lead to a qualitatively similar result.  

\begin{figure}[ht] 
\includegraphics[width=0.9\linewidth]{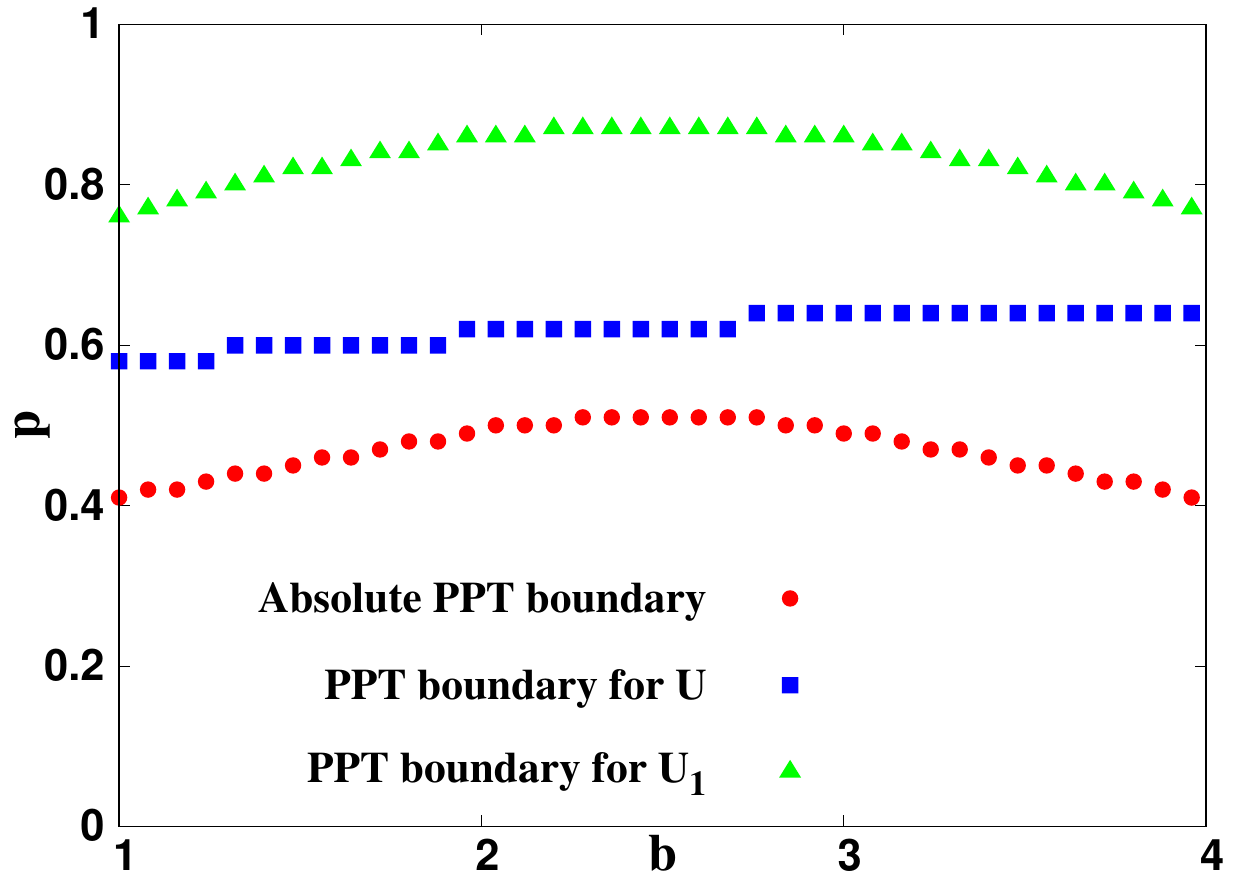}
\caption{The mixing parameter, \(p\), with respect to $b$ in the state, \(\rho_3\) in Eq. (\ref{eq:rho3}).  Here we choose \(U (\pi/18, 5\pi/6)\) and \(U_1\) as given in Eqs. (\ref{eq:U3x3}) and (\ref{eq1:U3x3}).   All other specifications are same as in Fig. \ref{fig:2}. 
}
\label{fig:4}
\end{figure}

\subsubsection{Detecting non-absolute PPT states in $3\otimes3$} 

Let us move to an example of a class of two-qutrit state. This example is different than the examples considered before since in $3\otimes3$, the set of absolutely separable states and absolutely PPT states are different and we concentrate on the detection of non-absolutely PPT states. 
In this purpose, let us consider a state 
\begin{equation}
\label{eq:rho3}
\rho_{3}=p\rho'_{b} +\dfrac{1-p}{9}I_{9\times9},
\end{equation}
where $\rho_{b}$ \cite{horo'99} is given by
\begin{equation}
    \rho'_b= \dfrac{2}{7}|\tilde{\psi}\rangle\langle\tilde{\psi}|+\dfrac{b}{7}\sigma_{+}+\dfrac{5-b}{7}\sigma_{-}.
    \end{equation}
Here $|\tilde{\psi}\rangle=\dfrac{1}{\sqrt{3}}(|00\rangle+|11\rangle+|22\rangle)$,  $\sigma_{+}=\dfrac{1}{3}(|01\rangle\langle01|+|12\rangle\langle12|+|20\rangle\langle20|)$, 
and 
$\sigma_{-}=\dfrac{1}{3}(|10\rangle\langle10|+|21\rangle\langle21|+|02\rangle\langle02|)$. 
The state $\rho'_b$ is PPT for $1 \leq{b}\leq  4$, and so we confine in this range of  $b$.
By using inequalities, (\ref{eq:absPPT1}) and (\ref{eq:absPPT2}),  
we provide the range of $b$ and $p$ for which the state $\rho_{3}$ is absolutely PPT (see the red circles in Fig. \ref{fig:4} for the boundary of absolutely PPT and non-absolutely PPT states).
Like Eq. (\ref{eq:U2x4}),  the unitary operator in this case reads as
\begin{eqnarray}
\label{eq:U3x3}
   U (\phi_1, \phi_2) &=& \cos \phi_1 [\sigma_x\otimes\sigma_y\otimes\sigma_z] \nonumber \\
   &+ & \sin \phi_1 \sin \phi_2 [\sigma_y\otimes\sigma_z\otimes\sigma_x]\nonumber \\
   & + & \sin \phi_1 \cos \phi_2 [\sigma_z\otimes\sigma_x\otimes\sigma_y] \oplus[1],
\end{eqnarray}
where $[1]$ is a $1\times1$ matrix with entry $1$.
A unitary operator  of this class, specifically,  $U(\pi/18, 5 \pi/6)$,  is capable to make non-absolutely PPT states to NPPT for some region in the  $(b,p)$-plane as depicted by squares in Fig. \ref{fig:4}. Notice that  we require different unitary operator if the non-absolutely PPT states belonging to the region between circles and squares have to
make NPPT (as also seen in Fig. \ref{fig:2}).


Like in the previous example, we fix  $ b=1.5$.  Let us first construct a linear witness operator, 
$W=U(\pi/18, 5 \pi/6)^\dagger|\phi\rangle\langle\phi|^{T_B}U(\pi/18, 5 \pi/6)$, which detects the state $\rho_{3}$ as non-absolutely PPT for $0.6 < p \leq 1$. In this case,  $|\phi\rangle = (-0.4476-0.004054i,\;-0.0103-0.009966i,\;-0.001158+0.3953i,\;0.02944-0.04832i,\;0.0003527-0.001285i,\;-0.05052+0.01027i,\;0.000449-0.3918i,\;-0.0478-0.03061i,\;0.6933)^T$. 

\begin{figure}[ht] 
\includegraphics[width=0.9\linewidth]{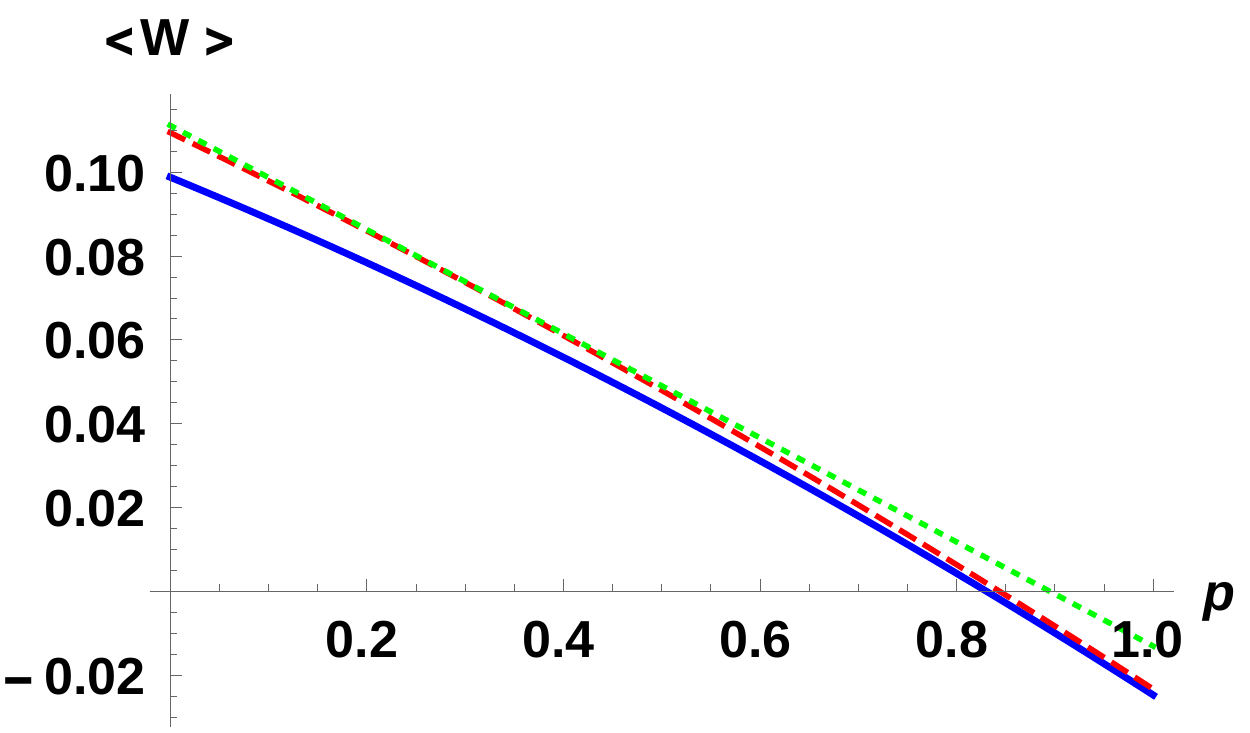}
\caption{Plots of nonlinear and linear witness operators, \(\langle W\rangle\), against the noise parameter $p$ for \(\rho_3\) in Eq. (\ref{eq:rho3}). Here $b =1.5$. The choices of \(\{p_i\}\)s and basis required for witness operators are mentioned in the text.  All other specifications are same as in Fig. \ref{fig:1}. 
}
\label{fig:5}
\end{figure}

To portray the power of nonlinear witness operator, let us choose  $|\phi'\rangle = (p_1+i p_2,\;-0.0103-0.009966i,\;-0.001158+0.3953i,\;0.02944-0.04832i,\;0.0003527-0.001285i,\;-0.05052+0.01027i,\;0.000449-0.3918i,\;-0.0478-0.03061i,\;p_3+ip_4)^T$,  with $p_1=-0.564882,\;p_2=0.471498,\;p_3=0.373546,\;p_4=0.0 $, and for \(F^1\) and \(F^2\), 
$|\psi\rangle=\frac{1+i}{\sqrt{2}}|22\rangle$  and the orthonormal basis as $\frac{1+i}{\sqrt{2}}\{|00\rangle,|01\rangle,|02\rangle,|10\rangle,|11\rangle,|12\rangle,|20\rangle,|21\rangle,|22\rangle\}$ respectively. Notice that \(|\phi'\rangle\) is  almost same as  \(|\phi\rangle\) in the linear witness operator except the first and the last entries.  
With these parameter values,  we  again identify a range of   the noise value, $p$, for which the performance of \(F^2\) is better than \(F^1\) and the linear witness operator (see Fig. \ref{fig:5}). 

\section{DETECTION LOOPHOLE IN NONLINEAR WITNESS operator for non-absolutely separable (PPT) states}
\label{sec:loophole}

Upto now, all the results derived in Secs. \ref{sec:twoqubits}  and \ref{sec:highdim} are under the assumption that the apparatuses used in the identification process are perfect. We will now  investigate the response of imperfect detectors like no clicking of detectors  (lost events) or additional events  on nonlinear witnesses \cite{dl'07, ks'19}. We find critical efficiency of detectors so as to avoid the fake detection of resources. 

Towards fulfilling the aim, we decompose a witness operator in local operator basis such that $W=C_{0} I+\sum_{\alpha} C_\alpha{S}_\alpha$, where I stands for the identity operator, $S_\alpha$ represents a term from expansion of $W$ in local operator basis and $C_\alpha$ is the corresponding expansion coefficient \cite{dl'07}. Using this decomposition, it can be shown that the condition for detecting a NPPT state using linear witness operator experimentally when detector does not work ideally is given by
\begin{equation}
    \langle{W}_\phi\rangle_{m}<C_{0}(1-\frac{1}{\eta_-})
\end{equation}
where $\langle{W}_\phi\rangle_{m}$ is experimentally measured value of the witness operator and $\eta_-$ is the lost event efficiency of detector (which is supplied). We here assume that the  additional event efficiency vanishes. On the other hand, in case of non-linear witness operator, $F^1$, the relation between the measured and the true values  modifies as \cite{ks'19}
\begin{equation}
\label{eqWnlineff}
\begin{split}
    \langle{W}_\phi\rangle_{m}<C_{0}(1-\frac{1}{\eta_-})+\frac{\eta_-}{S(\psi)}(\langle{H}\rangle_{m}^2+{K}_H^2-2\langle{H}\rangle_{m}K_H)\\+\frac{\eta_-}{S(\psi)}(\langle{A}\rangle_{m}^2+{K}_A^2-2\langle{A}\rangle_{m}K_A),
    \end{split}
\end{equation}
where $\langle{H}\rangle_m$ and $\langle{A}\rangle_m$ are experimentally measured value of Hermitian and anti-Hermitian part of $X^{T_B}$ respectively. Moreover $K_H$ and $K_A$ are given by
\begin{equation}
\label{eq:KHKA}
    K_H=C_{0H}(1-\frac{1}{\eta_-}) \text{   and   }  K_A=C_{0A}(1-\frac{1}{\eta_-}),
\end{equation}
with $C_{0H}$ and $C_{0A}$ being the coefficients of the decomposition of $H$ and $A$ in local operator basis respectively, corresponding to the Identity matrix.

It is possible to obtain \(  \langle{W}_\phi\rangle_{m}\) for all the examples that we considered so far. For example, when  $\eta_-$ is supplied,  for the generalized two-qubits Werner state,  $\langle{W}_\phi\rangle_{m}<\frac{1}{4}(1-\frac{1}{\eta_-})$
 with linear witness operator while for nonlinear case, $\langle{W}_\phi\rangle_{m}<\frac{1}{4}(1-\frac{1}{\eta_-})+\eta_{-}\{(\langle{H}\rangle_{m}-K_H)^{2}+\langle{A}\rangle_{m}^2\}$ with $K_H = \frac{1}{4}\cos \theta(1-\frac{1}{\eta_-})$ and $K_A =0$. 
So, for two-qubits, the loophole cannot be closed for our chosen linear witness (with $\theta=0.352$ ) if $\eta_-< 0.43582$ while  in case of nonlinear witness operator, for a given \(X_{nl} = 0.4\), the upper bound of \(\eta_{-}\) can further be lowered as \(0.3881\) where we assume that \(|\psi\rangle\) is orthogonal to \(|\phi \rangle\) in \(F^1\). 

Considering a qubit-qudit state \(\rho_2\)  given in Eq. (\ref{eq:rho2}), let us demonstrate the advantageous role of nonlinear witness operators towards defeating the inefficiency in detectors. In this case, the loophole of a linear witness can be closed when $\langle{W}_\phi\rangle_{m}<\frac{1}{8}(1-\frac{1}{\eta_-})$, while the similar condition in presence of nonlinearity reads as $\langle{W}_\phi\rangle_{m}<\frac{1}{8}(1-\frac{1}{\eta_-})+\eta_{-}\{(\langle{H}\rangle_{m}-K_H)^{2}+\langle{A}\rangle_{m}^2\}$, where $K_H = \frac{1}{8\sqrt{2}}(1-\frac{1}{\eta_-})\sin\theta_1\sin\theta_2\sin\theta_3(\cos\theta_4+\sin\theta_4\sin\theta_5\sin\theta_6\sin\theta_7)$ and \(K_A =0\).  

\begin{figure}[ht] 
	\includegraphics[width=0.9\linewidth]{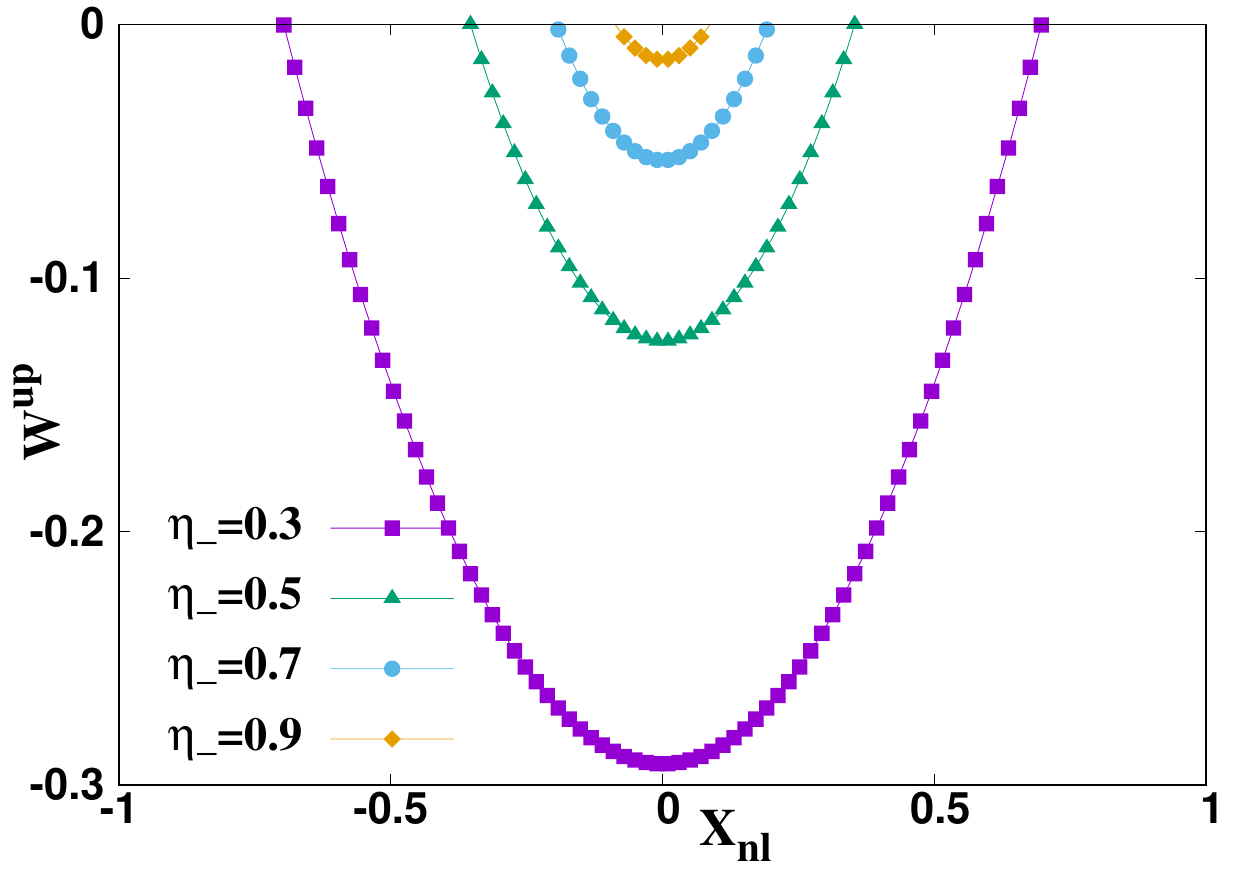}
	\caption{Inefficiency in detectors.  The upper bound of \(\langle W_{\phi}\rangle_m\) in the inequality (\ref{eq:noneffasol}), \(W^{up}\),  against  $X_{nl} = (\langle{H}\rangle_{m}^{2}+\langle{A}\rangle_{m}^2)$ for \(\rho_2\). We plot for different values of \(\eta_{-}\).
	\(|\phi\rangle\) chosen for computing \(F_1\) and \(U\) are same as in Fig. \ref{fig:3}. The behavior of \(W^{up}\) indicates that and with the increase of lost events inefficiency, \(\eta_{-}\), nonlinearity helps to obtain the high measured witness value. 
	Both the axes are dimensionless. }
	\label{fig:6}
\end{figure}

Let us consider a special case when $|\psi\rangle$ is orthogonal to $|\phi\rangle$, thereby leading to the vanishing \(K_H\) and \(K_A\).  For a fixed \(\eta_{-}\), the lost events inefficnecy can be overcome when
\begin{equation}
\label{eq:noneffasol}
  \langle{W}_\phi\rangle_{m}< \frac{1}{8}(1-\frac{1}{\eta_-})+\frac{\eta_{-}}{S(\psi)}\{\langle{H}\rangle_{m}^{2}+\langle{A}\rangle_{m}^2\} \equiv W^{up}.   
\end{equation}
The above equation can be rewritten if we consider $(\langle{H}\rangle_{m}^{2}+\langle{A}\rangle_{m}^2)=X_{nl}^2$. 
In Fig. \ref{fig:6},  the upper bound of \( \langle{W}_\phi\rangle_{m}\) is shown with respect to \(X_{nl}\) for different values of \(\eta_{-}\) which indicates that the chances of detecting non-absolutely separable states (with respect to the detector efficiency) increase with the increasing value of  nonlinearity, $X_{nl}^2$.
To visualize it, let us consider the situation when $\langle{W}_\phi\rangle_m$ vanishes and the nonlinear term is taken to be $X_{nl}=0.2$.  The detection of non-AS state   is  then possible when $\eta_-\geq0.424$, while for  $X_{nl}=0.6$, $\eta_-\geq 0.275$. Therefore, the increase of nonlinearity, i.e.,  $X_{nl}$ enhances the possibility of identifying non-AS states  even in presence of a relatively inferior detector. If we compare the values with the linear witness operators, we can also show that nonlinearity in the witness operator helps to overcome the detection loopholes. 



The similar detection inefficiency can also be overcome when the task is to identify the range of $p$ in \(\rho_3\) representing the non-absolute PPT states. 
In this case,  $K_H$ and $K_A$ can be evaluated to be $\frac{1}{9\sqrt{2}}(p_3+p_4)(1-\frac{1}{\eta_-})$ and $\frac{1}{9\sqrt{2}}(p_3-p_4)(1-\frac{1}{\eta_-})$ respectively and hence the condition in (\ref{eqWnlineff}) can also be obtained for a fixed \(\eta_-\).  


\section{Conclusion}
\label{sec:conclu} 

Quantum information processing tasks can only  successfully  be realized if the  resource  states required for that particular job are prepared and identified in an efficient manner. Among several available resources, entanglement shared between multiple parties has become one of the important ingredients in most of the quantum protocols discovered to date. Interestingly, however, it was found that there exists a set of unentangled states which can be made entangled by global unitary transformation while the rest  of the states remains useless, known as absolutely separable states. 

Therefore, in the development of quantum technologies, determining  non-absolutely separable states in laboratories can be a significant issue. Among several identification methods developed in the theory of entanglement, the most experimental-friendly one is the linear witness operators although, for a given state,  the general method of obtaining an optimal witness operator is still not known. On the other hand, nonlinear witness operators are shown to be  good alternatives  with respect to the detection of useful resources. 

The present work develops  nonlinear witness operators for detecting non-absolutely separable as well as non-absolutely positive partial transposed (PPT) states in presence of an ideal and inefficient detector. We explicitly constructed nonlinear witness operators for the class of two-qubit states  and admixtures of bound entangled states with white noise in different dimensions. Specifically, we showed that in presence of noise, nonlinear witness operators can be more efficient to recognize non-absolutely PPT states compared to their linear counterparts. Moreover, we found that such advantages are more pronounced when the detectors are inefficient.

\section*{Acknowledgement}
We acknowledge the support from Interdisciplinary Cyber Physical Systems (ICPS) program of the Department of Science and Technology (DST), India, Grant No.: DST/ICPS/QuST/Theme- 1/2019/23. SM acknowledges Ministry of Science and Technology in Taiwan (Grant no. 110-2811-M-006 -501).

\section*{Appendix: different unitary lead to different witness operator}
It is to be noted that different unitary transformations make  a PPT state to a NPPT one in a different parameter-range. For  example, motivated by the unitary transformation in two-qubits,    we write a global unitary operator in  $2\otimes4$ as 
\begin{equation}
\label{eq1:U2x4}
U_1 =\dfrac{1}{\sqrt{2}}
\begin{pmatrix}
1 & 0 & 0 & 0 & 0 & 0 & 0 & 1 \\ 0 & \sqrt{2} & 0 & 0 & 0 & 0 & 0 & 0 \\ 0 & 0 & \sqrt{2} & 0 & 0 & 0 & 0 & 0\\ 0 & 0 & 0 & \sqrt{2} & 0 & 0 & 0 & 0 \\ 0 & 0 & 0 & 0 & \sqrt{2} & 0 & 0 & 0 \\ 0 & 0 & 0 & 0 & 0 & \sqrt{2} & 0 & 0 \\ 0 & 0 & 0 & 0 & 0 & 0 & \sqrt{2} & 0 \\ -1 & 0 & 0 & 0 & 0 & 0 & 0 & 1 
\end{pmatrix}
\end{equation}
while for $3\otimes3$, a similar unitary transformation takes the form as
\begin{equation}
\label{eq1:U3x3}
U_1=\dfrac{1}{\sqrt{2}}\begin{pmatrix}
1 & 0 & 0 & 0 & 0 & 0 & 0 & 0 & 1 \\ 0 & \sqrt{2} & 0 & 0 & 0 & 0 & 0 & 0 & 0 \\ 0 & 0 & \sqrt{2} & 0 & 0 & 0 & 0 & 0 & 0\\ 0 & 0 & 0 & \sqrt{2} & 0 & 0 & 0 & 0 & 0 \\ 0 & 0 & 0 & 0 & \sqrt{2} & 0 & 0 & 0 & 0 \\ 0 & 0 & 0 & 0 & 0 & \sqrt{2} & 0 & 0 & 0 \\ 0 & 0 & 0 & 0 & 0 & 0 & \sqrt{2} & 0 & 0 \\ 0 & 0 & 0 & 0 & 0 & 0 & 0 & \sqrt{2} & 0 \\ -1 & 0 & 0 & 0 & 0 & 0 & 0 & 0 & 1	
\end{pmatrix}.
\end{equation}
Figs.  \ref{fig:2} and \ref{fig:4}, triangles symbolize the boundary between PPT states which can be made NPPT via these unitary operators and the PPT states which remain PPT even after their applications. Clearly, the above unitary operators are weaker than \(U\)  in Eqs. (\ref{eq:U2x4}) and (\ref{eq:U3x3}).  

\begin{thebibliography}{99}
\bibitem{dimexplain} \(d_1 \otimes d_2\) represents the dimensions of the complex Hilbert space, \(d_1 d_2\) where \(d_1\) and \(d_2\) are the dimensions of subsystems \(A\) and \(B\) respectively.
\end{thebibliography}

\end{document}